# Single-shot autofocusing of microscopy images using deep learning

Yilin Luo[1,2,3]†, Luzhe Huang[1,2,3]†, Yair Rivenson[1,2,3]*, Aydogan Ozcan[1,2,3,4]*

1 Electrical and Computer Engineering Department, University of California, Los Angeles, California 90095, USA

2 Bioengineering Department, University of California, Los Angeles, California 90095, USA

3 California Nano Systems Institute (CNSI), University of California, Los Angeles, California 90095, USA

4 David Geffen School of Medicine, University of California Los Angeles, California 90095, USA

* ozcan@ucla.edu, rivensonyair@ucla.edu

† Contributed equally

**ABSTRACT:** Autofocusing is a critical step for high-quality microscopic imaging of specimens, especially for measurements that extend over time covering large fields-of-view. Autofocusing is generally practiced using two main approaches. Hardware-based optical autofocusing methods rely on additional distance sensors that are integrated with a microscopy system. Algorithmic autofocusing methods, on the other hand, regularly require axial scanning through the sample volume, leading to longer imaging times, which might also introduce phototoxicity and photobleaching on the sample. Here, we demonstrate a deep learning-based *offline* autofocusing method, termed Deep-R, that is trained to rapidly and blindly autofocus a single-shot microscopy image of a specimen that is acquired at an arbitrary out-of-focus plane. We illustrate the efficacy of Deep-R using various tissue sections that were imaged using fluorescence and brightfield microscopy modalities and demonstrate snapshot autofocusing under different scenarios, such as a uniform axial defocus as well as a sample tilt within the field-of-view. Our results reveal that Deep-R is significantly faster when compared with standard online algorithmic autofocusing methods. This deep learning-based blind autofocusing framework opens up new opportunities for rapid microscopic imaging of large sample areas, also reducing the photon dose on the sample.

## Introduction

A critical step in microscopic imaging over an extended spatial or temporal scale is focusing. For example, during longitudinal imaging experiments, focus drifts can occur as a result of mechanical or thermal fluctuations of the microscope body[1] or microscopic specimen movement when for example live cells or model organisms are imaged. Another frequently encountered scenario which also requires autofocusing is due to the nonuniformity of the specimen's topography[2]. Manual focusing is impractical, especially for microscopic imaging over an extended period of time or a large specimen area.

Conventionally, microscopic autofocusing is performed "online", where the focus plane of each individual field-of-view (FOV) is found during the image acquisition process. Online autofocusing can be generally categorized into two groups: optical[3–9] and algorithmic methods[10–13]. Optical methods typically adopt additional distance sensors involving e.g., a near-infrared laser[3–5], a light-emitting diode[6] or an additional camera[7–9,14], that measure or calculate the relative sample distance needed for the correct focus. These optical methods require modifications to the optical imaging system, which are not always compatible with the existing microscope hardware[15]. Algorithmic methods, on the other hand, extract an image sharpness function/measure at different axial depths and locate the best focal plane using an iterative search algorithm. However, the focus function is in general sensitive to the image intensity and contrast, which in some cases can be trapped in a false local maxima/minima[16]. Another limitation of these algorithmic autofocusing methods is the requirement to capture multiple images through an axial scan (search) within the specimen volume. This process is naturally time-consuming, does not support high frame-rate imaging of dynamic specimen and increases the probability of sample photobleaching, photodamage or phototoxicity[17]. As an alternative, wavefront sensing-based autofocusing techniques[18–20] also lie at the intersection of optical and algorithmic methods. However, multiple image capture is still required, and therefore these methods also suffer from similar problems as the other algorithmic autofocusing methods face.

In recent years, deep learning has been demonstrated as a powerful tool in solving various inverse problems in microscopic imaging[21], for example, cross-modality super-resolution[22,23], virtual staining[24,25], localization microscopy[26,27], phase recovery and holographic image reconstruction[28–30]. Unlike most inverse problem solutions that require a carefully formulated forward model, deep learning instead uses image data to indirectly derive the relationship between the input and the target output distributions. Once trained, the neural network takes in a new sample's image (input) and rapidly reconstructs the desired output without any iterations, parameter tuning or user intervention.

Motivated by the success of deep learning-based solutions to inverse imaging problems, recent works have also explored the use of deep learning for online autofocusing of microscopy images[15,16,31,32]. Some of these previous approaches combined hardware modifications to the microscope design with a neural network; for example, Pinkard et al. designed a fully connected Fourier neural network (FCFNN) that utilized additional off-axis illumination sources to predict the axial focus distance from a single image[31]. As another example, Jiang et al. treated autofocusing as a regression task and employed a convolutional neural network (CNN) to estimate the focus distance without any axial scanning[15]. Dastidar et al. improved upon this idea and proposed to use the difference of two defocused images as input to the neural network, which showed higher focusing accuracy[16]. However, in the case of an uneven or tilted specimen in the FOV, all the techniques described above are unable to bring the whole region into focus simultaneously. Recently, a deep learning based virtual re-focusing method which can handle non-uniform and spatially-varying blurs has also been demonstrated[32]. By appending a pre-defined digital propagation matrix (DPM) to a blurred input image, a trained neural network can digitally refocus the input image onto a user-defined 3D surface that is mathematically determined by the DPM. This approach, however, does not perform autofocusing of an image as the DPM is user-defined, based on the specific plane or 3D surface that is desired at the network output.

Other post-processing methods have also been demonstrated to restore a sharply focused image from an acquired defocused image. One of the classical approaches that has been frequently used is to treat the defocused image as a convolution of the defocusing point spread function (PSF) with the in-focus image. Deconvolution techniques such as the Richardson-Lucy[33,34] algorithm require accurate prior knowledge of the defocusing PSF, which is not always available. Blind deconvolution methods[35,36] can also be used to restore images through the optimization of an objective function; but these methods are usually computationally costly, sensitive to image signal-to-noise ratio (SNR) and the choice of the hyperparameters used, and are in general not useful if the blur PSF is spatially varying. There are also some emerging methods that adopt deep learning for blind estimation of a space-variant PSF in optical microscopy[37].

Here we introduce a deep-learning based offline autofocusing method, termed Deep-R (Fig. 1), that enables the blind transformation of a single-shot defocused image into an in-focus image without any prior knowledge of the defocus distance, its direction, or the blur PSF, whether it is spatially varying or not. Compared to the existing body of autofocusing methods that have been used in optical microscopy, Deep-R is unique in a number of ways: (1) it does not require any hardware modifications to an existing microscope design; (2) it only needs a single image capture to infer and synthesize the in-focus image, enabling higher imaging throughput and reduced photon dose on the sample, without sacrificing the resolution; (3) its autofocusing is based on a data-driven, non-iterative image inference process that does not require prior knowledge of the forward imaging model or the defocus distance; and (4) it is broadly applicable to blindly autofocus spatially uniform and non-uninform defocused images, computationally extending the depth of field (DOF) of the imaging system.

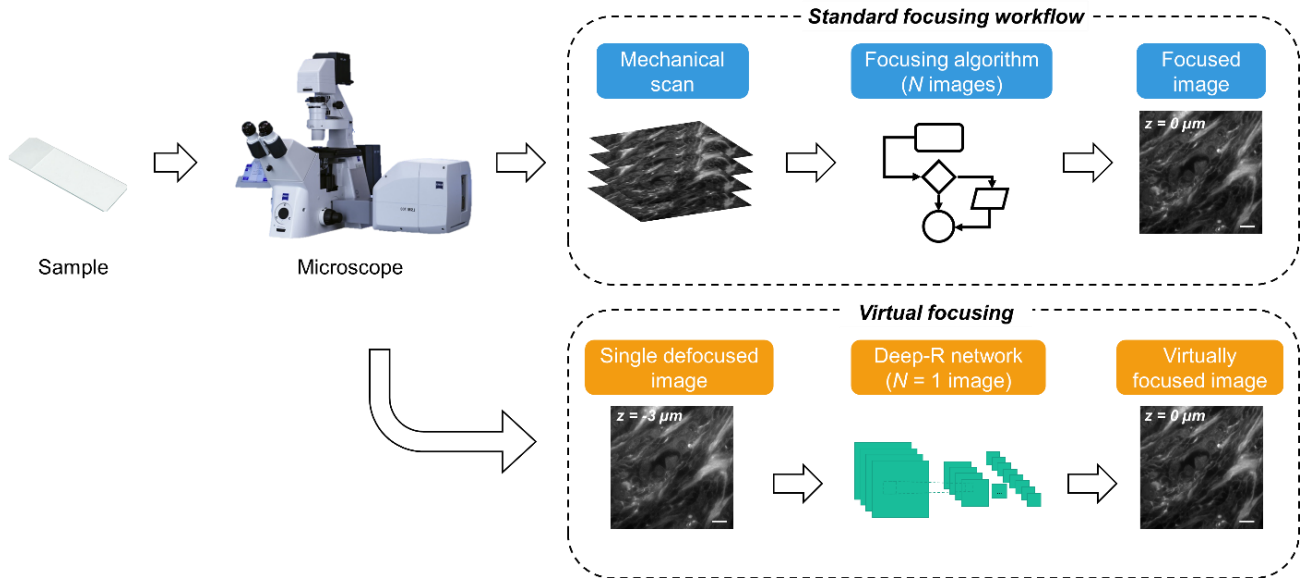

**Fig. 1. Deep-R autofocusing.** Deep-R blindly autofocuses a defocused image after its capture; mechanical autofocusing methods require multiple image acquisition at different axial locations.

Deep-R is based on a generative adversarial network (GAN)[38] that is trained with accurately paired in-focus and defocused image pairs. After its training, the generator network rapidly transforms a single defocused fluorescence image into an in-focus image. We demonstrated the performance of Deep-R using various fluorescence (including autofluorescence and immunofluorescence) and brightfield microscopy images with spatially uniform as well as non-uniform defocus within the FOV. Our results reveal that a trained Deep-R network significantly enhances the imaging speed of a benchtop microscope by ~15-fold by eliminating the need for axial scanning during the autofocusing process.

This data-driven offline autofocusing approach will be especially useful in high-throughput imaging over large sample areas, where focusing errors inevitably occur, especially over longitudinal imaging experiments. With Deep-R, the DOF of the microscope and the range of usable images can be significantly extended, thus reducing the time, cost and labor required for reimaging of out-of-focus areas of a sample. Simple to implement and purely computational, Deep-R can be applicable to a wide range of microscopic imaging modalities, as it requires no hardware modifications to the imaging system.

## Results and Discussion

### Deep-R based autofocusing of defocused fluorescence images

Fig. 2(a) demonstrates Deep-R based autofocusing of a single defocused immunofluorescence image of an ovarian tissue section. In the training stage, the network was fed with accurately paired/registered image data composed of (1) fluorescence images acquired at different axial defocus distances, and (2) the corresponding in-focus images (ground-truth labels), which were algorithmically calculated using an axial image stack ($N$ = 101 images captured at different planes; see the Methods section and Table S1). During the inference process, a pretrained Deep-R network blindly takes in a single defocused image at an arbitrary defocus distance (within the axial range included in the training), and digitally autofocuses it to match the ground truth image. Fig. 2(b) highlights a sample region of interest (ROI) to illustrate the blind output of Deep-R at different input defocus depths. Within the ± 5 μm axial training range, Deep-R successfully autofocuses the input images and brings back sharp structural details, e.g., corresponding to SSIM (structural similarity index) values above 0.7, whereas the mechanically scanned input images degrade rapidly, as expected, when the defocus distance exceeds ~0.65 μm, which corresponds to the DOF of the objective lens (40×/0.95NA). Even beyond its axial training range, Deep-R output images still exhibit some refocused features, as illustrated in Fig. 2(b).

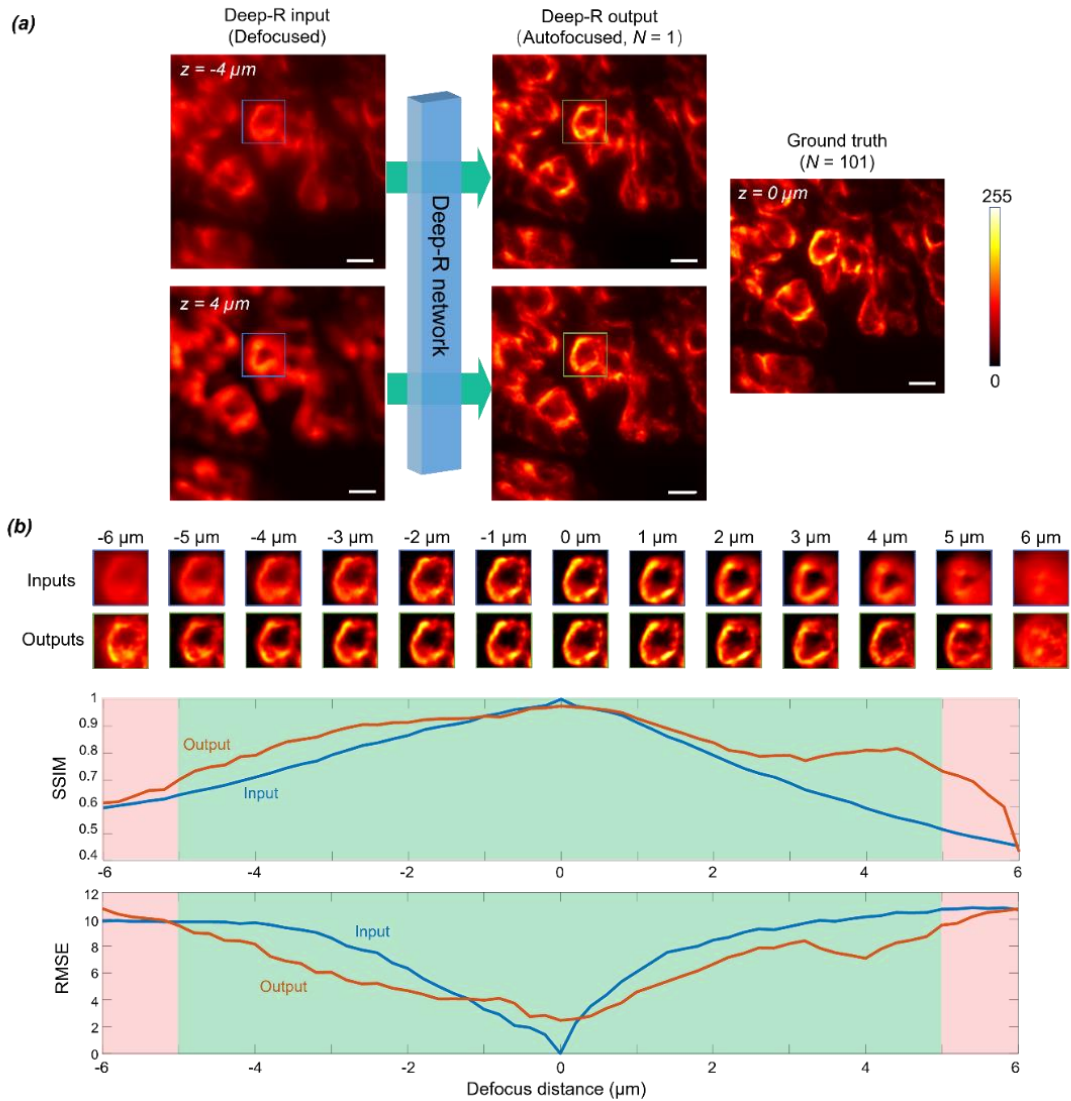

**Fig. 2. Deep-R based autofocusing of fluorescently stained samples.** (a) Deep-R performs blind autofocusing of individual fluorescence images without prior knowledge of their defocus distances or directions. Scale bars, 10 μm. (b) For the specific ROI in (a), the SSIM and RMSE values of input and output images with respect to the ground truth (z = 0 μm, in-focus image) are plotted as a function of the axial defocus distance. Green zone indicates that the axial defocus distance is within the training range while the red zone indicates that the axial range is outside of the training defocus range.

Similar blind inference results were also obtained for a densely-connected human breast tissue sample (see Fig. 3) that is imaged under a 20×/0.75NA objective lens, where Deep-R accurately autofocused the autofluorescence images of the sample within an axial defocus range of ± 5 μm.

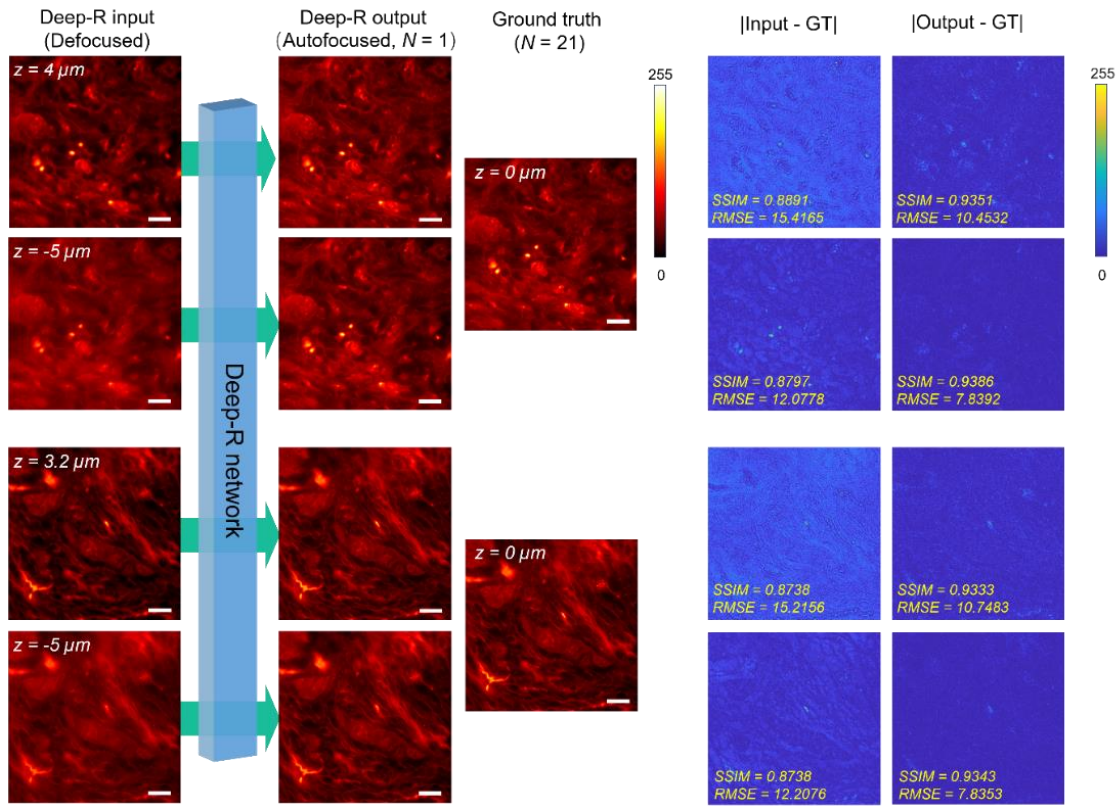

**Fig. 3. Deep-R based autofocusing of autofluorescence images.** Two different ROIs, each with positive and negative defocus distances, are blindly brought to focus by Deep-R. The absolute difference images of the ground truth with respect to Deep-R input and output images are also shown on the right, with the corresponding SSIM and RMSE quantification reported as insets. Scale bars, 20 μm.

**Point spread function analysis of Deep-R performance**

To further quantify the autofocusing capability of Deep-R, we imaged samples containing 300 nm polystyrene beads (excitation and emission wavelengths of 538 nm and 584 nm, respectively) using a 40×/0.95NA objective lens and trained two different neural networks with an axial defocus range of ± 5 μm and ± 8 μm, respectively. After the training phase, we then measured the 3D PSF of the input image stack and the corresponding Deep-R output image stack by tracking 164 isolated nanobeads across the sample FOV as a function of the defocus distance. For example, Fig. 4(a) illustrates the 3D PSF corresponding to a single nanobead, measured through this axial image stack (input images). As expected, this input 3D PSF shows increased spreading away from the focal plane. On the other hand, the Deep-R PSF corresponding to the output image stack of the same particle maintains a tighter focus, covering an extended depth, determined by the axial training range of the Deep-R network (see Fig. 4(a)). As an example, at z = -7 μm, the output images of a Deep-R network that is trained with ± 5 μm defocus range exhibit slight defocusing (see Fig. 4(b)), as expected. However, using a Deep-R network trained with ± 8 μm defocus range results in accurate refocusing for the same input images (Fig. 4(b)). Similar conclusions were observed for the blind testing of a 3D sample, where the nanobeads were dispersed within a volume spanning ~ 20 μm thickness (see Fig. S3).

Fig. 4(b) further presents the mean and standard deviation of the lateral full width at half maximum (FWHM) values as a function of the axial defocus distance, calculated from 164 individual nanobeads. The enhanced DOF of Deep-R output is clearly illustrated in the nearly constant lateral FHWM within the training range. On the other hand, the mechanically scanned input images show much shallower DOF, as reflected by the rapid change in the lateral FWHM as the defocus distance varies. Note also that the FWHM curve for the input image is unstable at the positive defocus distances, which is caused by the strong side lobes induced by out-of-focus lens aberrations. Deep-R output images, on the other hand, are immune to these defocusing introduced aberrations since it blindly autofocuses the image at its output and therefore maintains a sharp PSF across the entire axial defocus range that lies within its training, as demonstrated in Fig. 4(b).

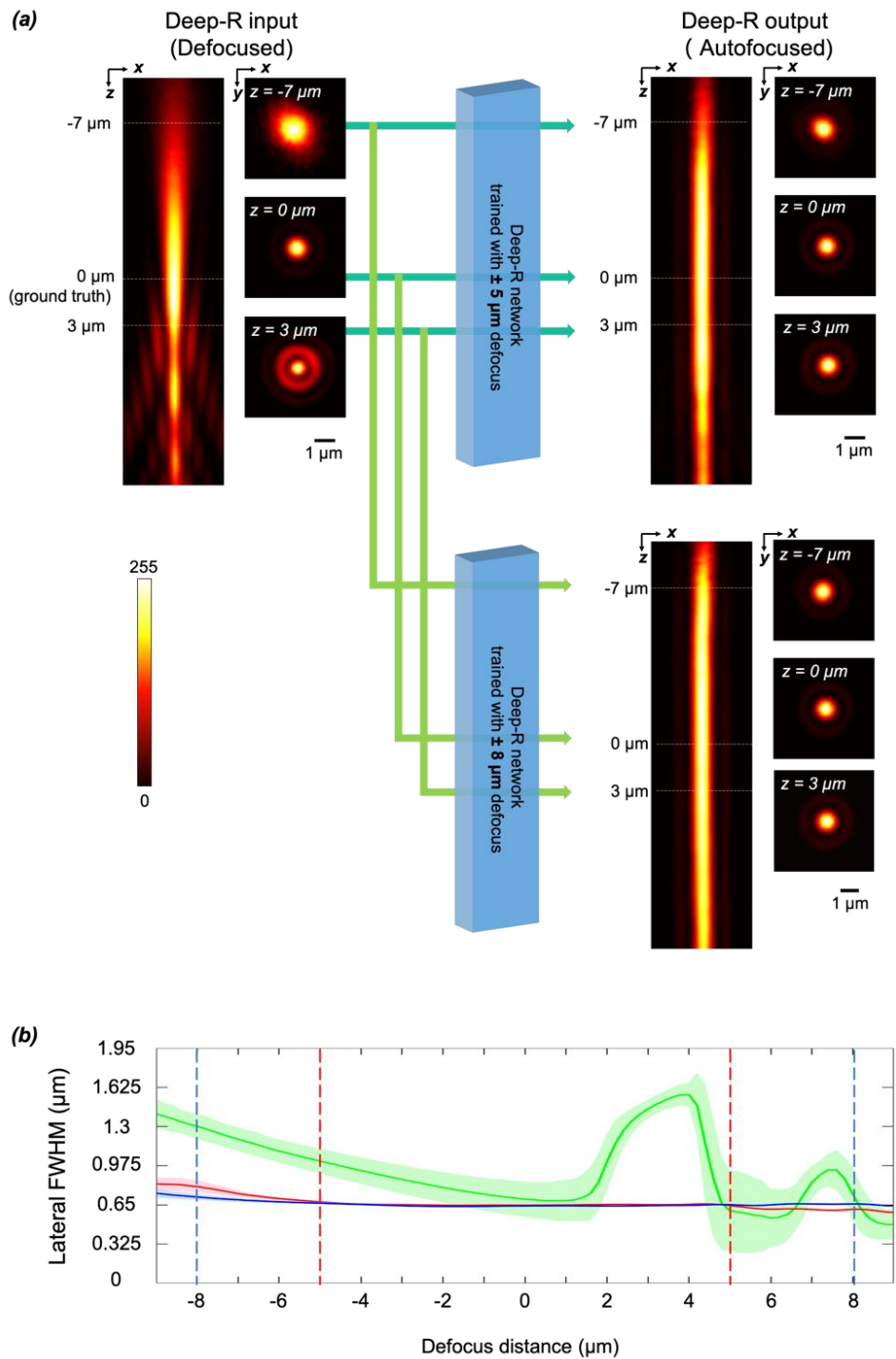

**Fig. 4. 3D PSF analysis of Deep-R using 300 nm fluorescent beads.** (a) Each plane in the input image stack is fed into Deep-R network and blindly autofocused. (b) Mean and standard deviations of the lateral FHWM values of the particle images are reported as a function of the axial defocus distance. The statistics are calculated from *N = 164* individual nanobeads. Green curve: FWHM statistics of the mechanically scanned image stack (i.e., the network input). Red curve: FWHM statistics of the output images calculated using a Deep-R network that is trained with ± 5 μm axial defocus range. Blue curve: FWHM statistics of the output images calculated using a Deep-R network that is trained with ± 8 μm axial defocus range.

**Comparison of Deep-R computation time against online algorithmic autofocusing methods**

In addition to post-correction of out-of-focus or aberrated images, Deep-R also provides a better alternative to existing online focusing methods, achieving higher imaging speed. Software-based conventional online autofocusing methods acquire *multiple* images at each FOV. The microscope captures the first image at an initial position, calculates an image sharpness feature, and moves to the next axial position based on a focus search algorithm. This iteration continues until the image satisfies a sharpness metric. As a result, the focusing time is prolonged, which leads to increased photon flux on the sample, potentially introducing photobleaching, phototoxicity or photodamage. This iterative autofocusing routine also compromises the effective frame rate of the imaging system, which limits the observable features in a dynamic specimen. In contrast, Deep-R performs autofocusing with a single-shot image, without the need for additional image exposures or sample stage movements, retaining the maximum frame rate of the imaging system. To better demonstrate this and emphasize the advantages of Deep-R, we experimentally measured the autofocusing time of 4 commonly used online focusing methods: Vollath-4 (VOL4) [39], Vollath-5 (VOL5)[39], standard deviation (STD) and normalized variance (NVAR) [10]. Table S2 summarizes our results, where we report the autofocusing time per 1mm$^2$ of sample FOV. Overall, these online algorithms take ~40 s/mm$^2$ to autofocus an image using a 3.5 GHz Intel Xeon E5-1650 CPU, while Deep-R inference takes ~ 20 s/mm$^2$ on the same CPU, and ~3 s/mm$^2$ on a Nvidia GeForce RTX 2080Ti GPU.

**Comparison of Deep-R autofocusing quality with offline deconvolution techniques**

Next, we compared Deep-R autofocusing against standard deconvolution techniques, specifically, the Landweber deconvolution[40] and the Richardson-Lucy (RL) deconvolution[33,34], using the ImageJ plugin DeconvolutionLab2[41]. For these offline deconvolution techniques, the lateral PSFs at the corresponding defocus distances were specifically provided using measurement data, since this information is required for both algorithms to approximate the forward imaging model. In addition to this a priori PSF information at different defocusing distances, the parameters of each algorithm were adjusted/optimized such that the reconstruction had the best visual quality for a fair comparison (see the Methods section). Figure 5 illustrates that at negative defocus distances (e.g., z = -3 μm), these offline deconvolution algorithms demonstrate an acceptable image quality in most regions of the sample, which is expected, as the input image maintains most of the original features at this defocus direction; however, compared with Deep-R output, the Landweber and RL deconvolution results showed inferior performance (despite using the PSF at each defocus distance as a priori information). A more substantial difference between Deep-R output and these offline deconvolution methods is observed when the input image is positively defocused (see e.g., z = 4 μm in Fig. 5). Deep-R performs improved autofocusing without the need for any direct PSF measurement or parameter tuning, which is also confirmed by the SSIM and RMSE (root mean square error) metrics reported in Fig. 5.

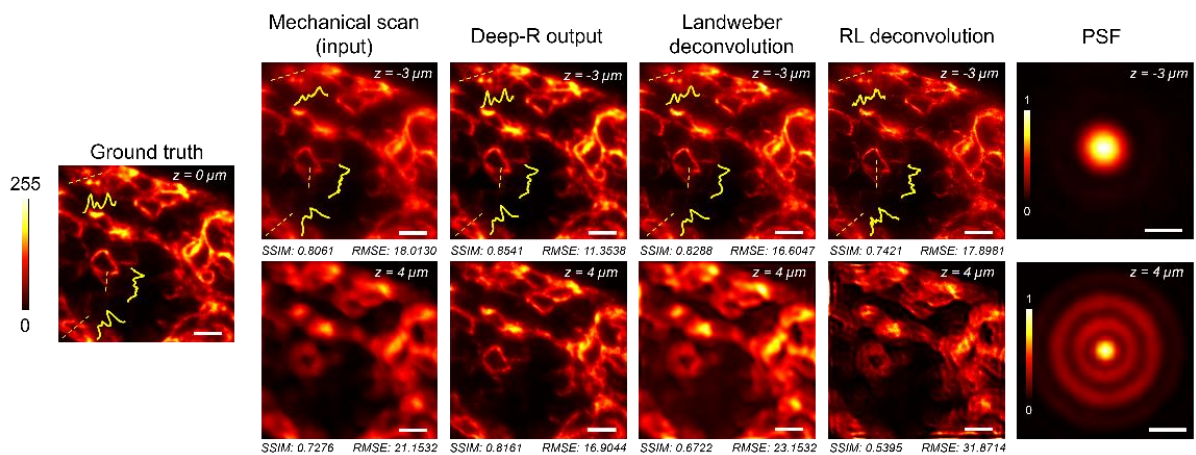

**Fig. 5. Comparison of Deep-R autofocusing with deconvolution techniques[33,34,40].** The lateral PSFs at the corresponding defocus distances are provided to the deconvolution algorithms as prior knowledge of the defocus model; Deep-R did not make use of the measured PSF information shown on the far right column; its inference model inherently "learned" that information during its training phase through image data. Scale bars for tissue images, 10 μm. Scale bars for PSF images, 1 μm.

**Deep-R based autofocusing of brightfield microscopy images**

While all the previous results are based on images obtained by fluorescence microscopy, Deep-R can also be applied to other incoherent imaging modalities, such as brightfield microscopy. As an example, we applied the Deep-R framework on brightfield microscopy images of an H&E (hematoxylin and eosin) stained human prostate tissue (Fig. 6). The training data were composed of images with an axial defocus range of ± 10 μm, which were captured by a 20×/0.75NA objective lens. After the training phase, the Deep-R network, as before, takes in an image at an arbitrary (and unknown) defocus distance and blindly outputs an in-focus image that matches the ground truth. Although the training images were acquired from a non-lesion prostate tissue sample, blind testing images were obtained from a different sample slide coming from a different patient, which contained tumor, still achieving high RMSE and SSIM accuracy at the network output (see Fig. 6 and Fig. S4), which indicates the generalization success of our presented method. The application of Deep-R to brightfield microscopy can significantly accelerate whole slide imaging (WSI) systems used in pathology by capturing only a single image at each scanning position within a large sample FOV, thus enabling high-throughput histology imaging.

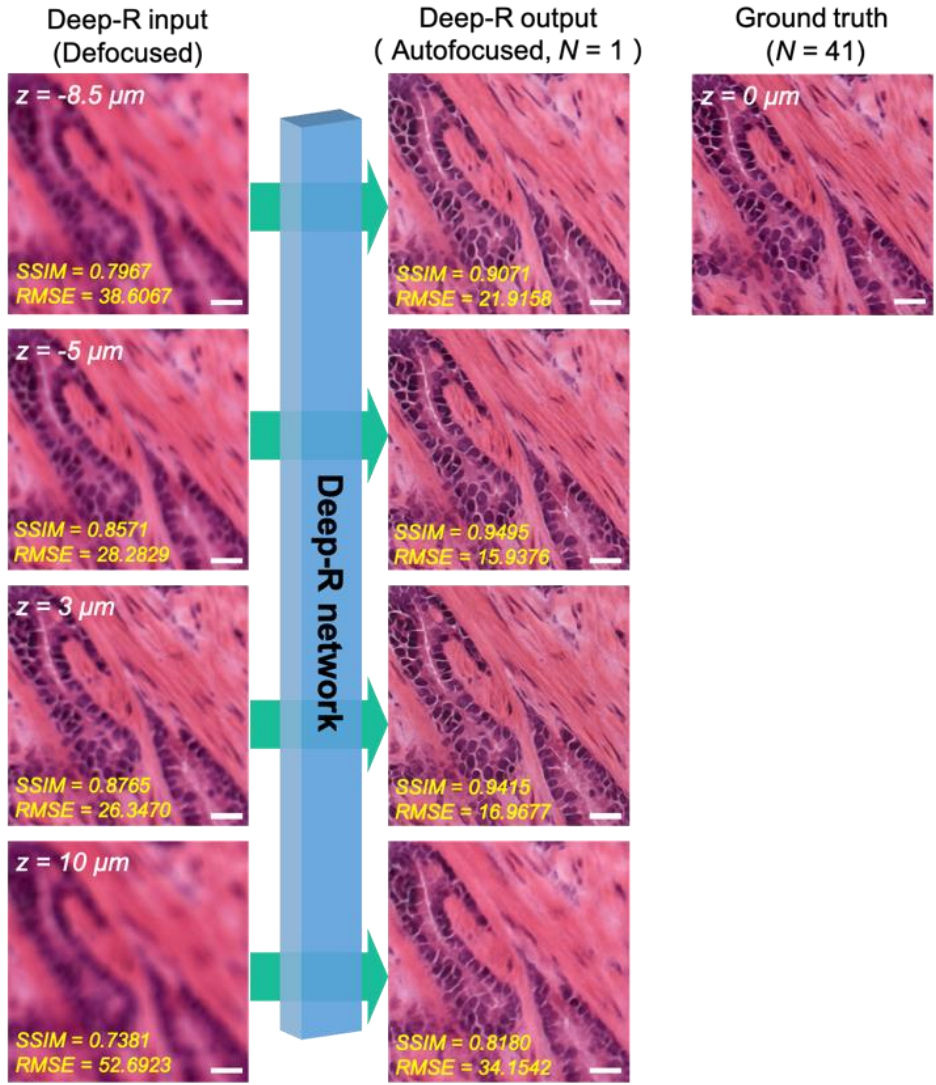

**Fig. 6. Deep-R based autofocusing of brightfield microscopy images.** The success of Deep-R is demonstrated by blindly autofocusing various defocused brightfield microscopy images of human prostate tissue sections. Scale bars, 20 μm.

**Deep-R autofocusing on non-uniformly defocused samples**

Next, we demonstrate that the axial defocus distance of every pixel in the input image is in fact encoded and can be inferred during Deep-R based autofocusing in the form of a DPM, revealing pixel-by-pixel the defocus distance of the input image (see Fig. 7). For this, a Deep-R network was first pre-trained without the decoder, following the same process as all the other Deep-R networks, and then the parameters of Deep-R were fixed. Next, a separate decoder with the same structure of the up-sampling path of the Deep-R network was separately optimized (see the Methods section) to learn the defocus DPM of an input image. In this optimization/learning process, only uniformly defocused images were used, i.e., the decoder was solely trained on uniform DPMs. Then, the decoder, along with the corresponding Deep-R, were both tested on uniformly defocused samples; as shown in Fig. 7(b), the output DPM matches the ground truth very well, successfully estimating the axial defocus distance of every pixel in the input image. As a further challenge, despite being trained using only *uniformly* defocused samples, the decoder was also blindly tested on a tilted sample with a tilt angle of 1.5°, and as presented in Fig. 7(c), the output DPM clearly revealed an axial gradient, corresponding to the tilted sample plane, demonstrating the generalization of the decoder to non-uniformly defocused samples.

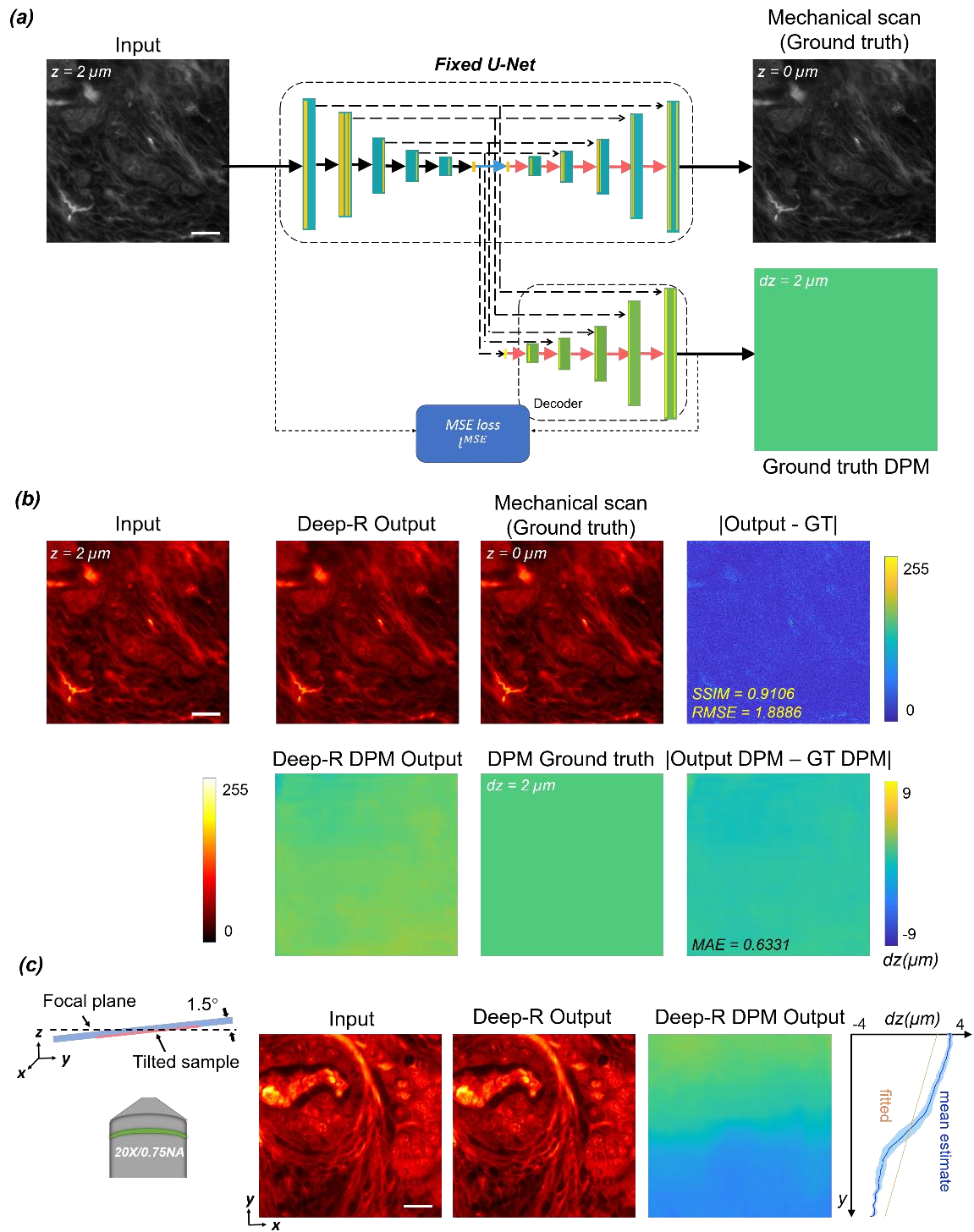

**Fig. 7. Pixel-by-pixel defocus distance extracted from an input image, in the form of a DPM.** (a) The decoder that is used to extract defocus distances from Deep-R autofocusing. The Deep-R network is pre-trained and fixed, and then a decoder is separately optimized to learn the pixel-by-pixel defocus distance in the form of a matrix, DPM. (b) The Deep-R autofocusing output and the extracted DPM on a uniformly defocused sample. (c) The Deep-R autofocusing output and the extracted DPM for a tilted sample. The $dz$-$y$ plot is calculated from the extracted DPM. Blue solid line: the mean $dz$ averaged by each row; Blue shadow: the standard deviation of the estimated $dz$ in each row; Yellow line: the fitted $dz$-$y$ line with a fixed slope corresponding to the tilt angle of the sample.

Next, Deep-R was further tested on non-uniformly defocused images that were this time generated using a pre-trained Deep-Z network fed with various non-uniform DPMs that represent tilted, cylindrical and spherical surfaces (Fig. 8).

Although Deep-R was exclusively trained on uniformly defocused image data, it can handle complex non-uniform defocusing profiles within a large defocusing range, with a search complexity of $O(1)$, successfully autofocusing each one of these non-uniformly defocused images shown in Fig. 8 in a single blind inference event. Furthermore, Deep-R autofocusing performance was also demonstrated using tilted tissue samples, detailed in Supplementary Section S.1. As illustrated in Fig. S1, at different focal depths (e.g., $z = 0$ µm and $z = -2.2$ µm), because of the tissue sample tilt, different sub-regions within the FOV were defocused by different amounts, but they were simultaneously autofocused by Deep-R, all in parallel, generating an extended DOF image that matches the reference fluorescence image.

**Generalization of Deep-R**

Although the blind autofocusing range of Deep-R can be increased by incorporating images that cover a larger defocusing range, there is a tradeoff between the inference image quality and the axial autofocusing range. To illustrate this tradeoff, we trained 3 different Deep-R networks on the same immunofluorescence image dataset as in Fig. 2, each with a different axial defocus training range, i.e., ± 2µm, ± 5µm, and ± 10µm, respectively. Fig. S5 reports the average and the standard deviation of RMSE and SSIM values of Deep-R input and output images, calculated from a blind testing dataset consisting of 26 FOVs, each with 512×512 pixels. As the axial training range increases, Deep-R accordingly extends its autofocusing range, as shown in Fig. S5. However, a Deep-R network trained with a large defocus distance (e.g., ± 10µm) partially compromises the autofocusing results corresponding to a slightly defocused image (see e.g., the defocus distances 2-5 µm reported in Fig. S5). Stated differently the blind autofocusing task for the network becomes more complicated when the axial training range increases, yielding a sub-optimal convergence for Deep-R (also see Fig. S6). A possible explanation for this behavior is that as the defocusing range increases, each pixel in the defocused image is receiving contributions from an increasing number of neighboring object features, which renders the inverse problem of remapping these features back to their original locations more challenging. Therefore, the inference quality and the success of autofocusing is empirically related to the sample density as well as the SNR of the acquired raw image.

As generalization is still an open challenge in machine learning, we wish to explore the generalization capabilities of our network in autofocusing images of new sample types that were not present during the training phase. For that, we used the public image dataset BBBC006v1 from the Broad Bioimage Benchmark Collection[42,43]. The dataset was composed of 768 image z-stacks of human U2OS cells, obtained using a 20× objective scanned using ImageXpress Micro automated cellular imaging system (Molecular Devices, Sunnyvale, CA). at two different channels for nuclei (Hoechst 33342, Ex/Em 350/461 nm) and phalloidin (Alexa Fluor 594 phalloidin, Ex/Em 581/609 nm), respectively, as shown in Supplementary Fig. S7(a). We separately trained 3 Deep-R networks with a defocus range of ± 10 µm on datasets that contain images of (1) only nuclei, (2) only phalloidin and (3) both nuclei and phalloidin, and tested their performance on images from different types of sample. As expected, the network has its optimal blind inference achieved on the same type of samples that it was trained with (Supplementary Fig. S7(b), yellow curves). Training with the mixed sample also generates similar results, with slightly higher RMSE error (Fig. S7(b), red curves). Interestingly, even when tested on images of a different sample type and wavelengths, Deep-R still performs autofocusing over the entire defocus training range (Fig. S7(b), green curves). A more concrete example is given in Fig. S7(c), where Deep-R is trained on the simple, sparse nuclei images, and still brings back some details when blindly tested on the densely connected phalloidin images.

One general concern for the applications of deep learning methods to microscopy is the potential generation of spatial artifacts and hallucinations. There are several strategies that we implemented to mitigate such spatial artifacts in output images generated by Deep-R. First, the statistics of the training process was closely monitored, by evaluating e.g., the validation loss and other statistical distances of the output data with respect to the ground truth images. As shown in Fig. S8, the training loss and validation loss curves demonstrate that a good balance, as expected, between the generator and discriminator networks was achieved and possible overfitting was avoided. Second, image datasets with sufficient structural variations and diversity were used for training. For example, ~1000 FOVs were included in the training datasets of each type of sample, covering 100 to 700 mm$^2$ of unique sample area (also see Table S1); each FOV contains a stack of defocused images from a large axial range (2 to 10 µm, corresponding to 2.5 to 15 times of the native DOF of the objective lens), all of which provided an input dataset distribution with sufficient complexity as well as an abstract mapping to the output data distribution for the generator to learn from. Third, standard practices in deep learning such as early stopping were applied to prevent overfitting in training Deep-R, as further illustrated in the training curves shown in Fig. S8. Finally, we should also note that when testing a Deep-R model on a new microscopy system that is different from the imaging hardware/configuration used in the training, it is generally recommended to either use a form of transfer learning with some new training image data acquired using the new microscopy hardware[45] or alternatively train a new model from scratch using new samples. Similarly, to use the presented approach on new types of samples that were not part of the training, fresh application of our framework or transfer learning are recommended, starting with the image registration between the input images and the desired labels corresponding to the new sample type.

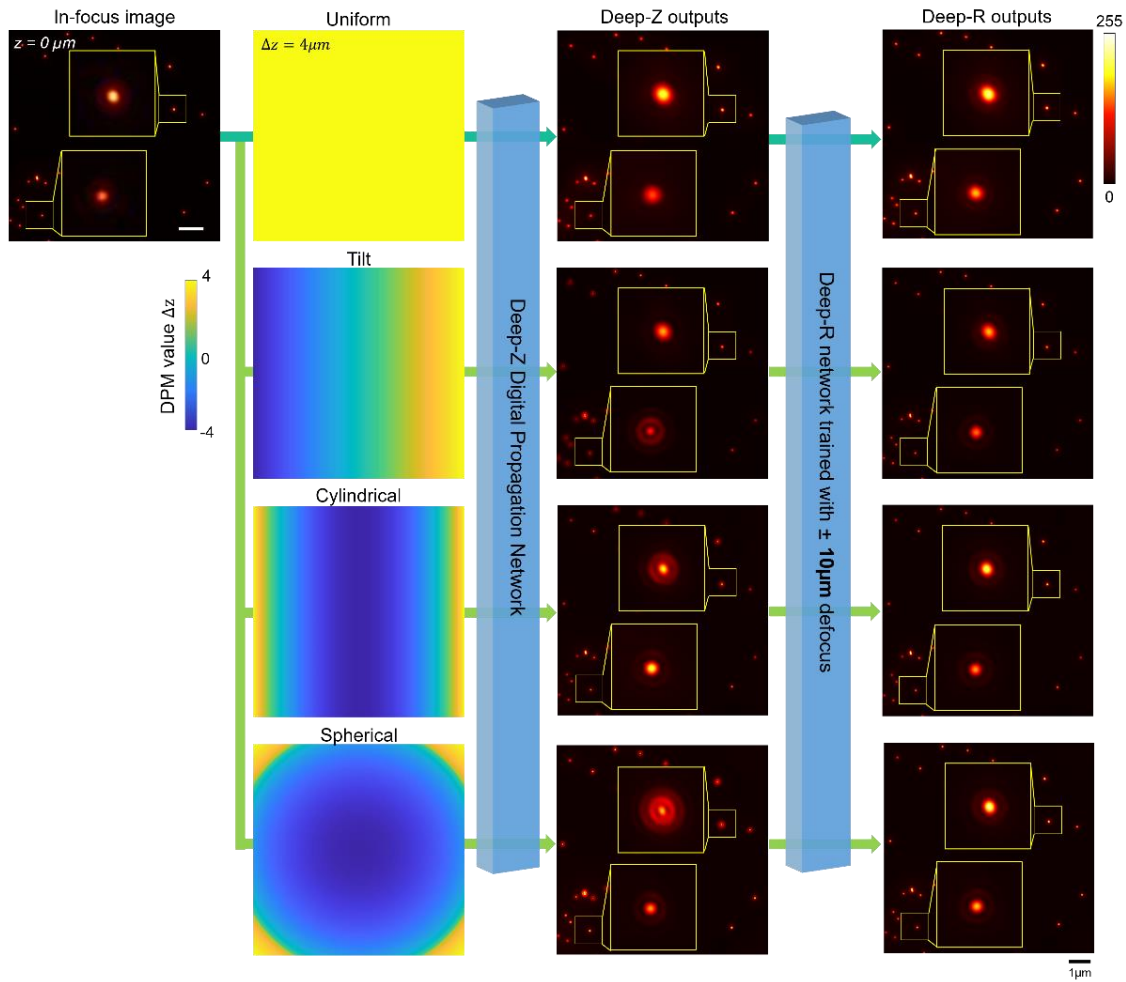

**Fig. 8 Deep-R autofocusing on non-uniformly defocused samples.** The non-uniformly defocused images were created by Deep-Z, using DPMs that represent tilted, cylindrical and spherical surfaces.

In summary, we presented a deep learning-based autofocusing framework, termed Deep-R, that enables offline, blind autofocusing from a single microscopy image. Although trained with uniformly defocused images, Deep-R can successfully autofocus images of samples that have non-uniform aberrations, computationally extending the DOF of the microscopic imaging system. This method is widely applicable to various incoherent imaging modalities e.g., fluorescence microscopy, brightfield microscopy and darkfield microscopy, where the inverse autofocusing solution can be efficiently learned by a deep neural network through image data. This approach significantly increases the overall imaging speed, and would especially be important for high-throughput imaging of large sample areas over extended periods of time, making it feasible to use out-of-focus images without the need for re-imaging the sample, also reducing the overall photon dose on the sample.

## Materials and Methods

### Sample preparation

*Breast, ovarian and prostate tissue samples:* the samples were obtained from the Translational Pathology Core Laboratory (TPCL) and prepared by the Histology Lab at UCLA. All the samples were obtained after the de-identification of the patient related information and prepared from existing specimens. Therefore, this work did not interfere with standard practices of care or sample collection procedures. The human tissue blocks were sectioned using a microtome into 4 μm thick sections, followed by deparaffinization using Xylene and mounting on a standard glass slide using CytosealTM (Thermo-Fisher Scientific, Waltham, MA, USA). The ovarian tissue slides were labelled by pan-cytokeratin tagged by fluorophore Opal 690, and the prostate tissue slides were stained with H&E.

*Nano-bead sample preparation:* 300 nm fluorescence polystyrene latex beads (with excitation/emission at 538/584nm) were purchased from MagSphere (PSFR300NM), diluted 3,000× using methanol. The solution is ultrasonicated for 20 min before and after dilution to break down clusters. 2.5 μL of diluted bead solution was pipetted onto a thoroughly cleaned #1 coverslip and let dry.

*3D nanobead sample preparation:* following a similar procedure as described above, nanobeads were diluted 3,000× using methanol. 10 μL of Prolong Gold Antifade reagent with DAPI (ThermoFisher P-36931) was pipetted onto a thoroughly cleaned glass slide. A droplet of 2.5 μL of diluted bead solution was added to Prolong Gold reagent and mixed thoroughly. Finally, a cleaned coverslip was applied to the slide and let dry.

### Image acquisition

The autofluorescence images of breast tissue sections were obtained by an inverted microscope (IX83, Olympus), controlled by the Micro-Manager microscope automation software. The unstained tissue was excited near the ultraviolet range and imaged using a DAPI filter cube (OSF13-DAPI-5060C, EX377/50, EM447/60, DM409, Semrock). The images were acquired with a 20×/0.75NA objective lens (Olympus UPLSAPO 20×/0.75NA, WD 0.65). At each FOV of the sample, autofocusing was algorithmically performed, and the resulting plane was set as the initial position (i.e., reference point). The autofocusing was controlled by the OughtaFocus plugin[44] in Micro-Manager[45], which uses Brent's algorithm[46] for searching of the optimal focus based on Vollath-5[39] criterion. For the training and validation datasets, the z-stack was taken from -10 μm to 10 μm with 0.5 μm axial spacing (DOF = 0.8 μm). For the testing image dataset, the axial spacing was 0.2 μm. Each image was captured with a scientific CMOS image sensor (ORCA-flash4.0 v.2, Hamamatsu Photonics) with an exposure time of ~100 ms.

The immunofluorescence images of human ovarian samples were imaged on the same platform with a 40×/0.95NA objective lens (Olympus UPLSAPO 40×/0.95NA, WD 0.18), using a Cy5 filter cube (CY5-4040C-OFX, EX628/40, EM692/40, DM660, Semrock). After performing the autofocusing, a z-stack was obtained from -10 μm to 10 μm with 0.2 μm axial steps.

Similarly, the nanobeads sample were imaged with the same 40×/0.95NA objective lens, using a Texas red filter cube (OSFI3-TXRED-4040C, EX562/40, EM624/40, DM593, Semrock), and a z-stack was obtained from -10 μm to 10 μm with 0.2 μm axial steps after the autofocusing step.

Finally, the H&E stained prostate samples were imaged on the same platform using brightfield mode with a 20×/0.75NA objective lens (Olympus UPLSAPO 20×/0.75NA, WD 0.65). After performing autofocusing on the automation software, a z-stack was obtained from -10 μm to 10 μm with an axial step size of 0.5 μm.

### Data pre-processing

To correct for rigid shifts and rotations resulting from the microscope stage, the image stacks were first aligned using the ImageJ plugin 'StackReg'[47]. Then, an extended DOF (EDOF) image was generated using the ImageJ plugin 'Extended Depth of Field'[48] for each FOV, which typically took ~180 s/FOV on a computer with i9-7900X CPU and 64GB RAM. The stacks and the corresponding EDOF images were cropped into non-overlapping 512×512-pixel image patches in the lateral direction, and the ground truth image (z = 0 μm) was set to be the one with the highest SSIM with respect to the EDOF image. Then, a series of defocused planes, above and below the focused plane, were selected as input images and input-label image pairs were generated for network training. The image datasets were randomly divided into training and validation datasets with a preset ratio of 0.85:0.15 with *no overlap* in FOV. Note also that the blind testing dataset was cropped from separate FOVs from *different* sample slides that did *not* appear in the training and validation datasets. Training images are augmented 8 times by random flipping and rotations during the training, while the validation dataset was not augmented. Each pair of input and ground truth images were normalized such that they have zero mean and unit variance before they were fed into the corresponding Deep-R network. All FOVs were captured with ~15% spatial overlap, and the total unique sample area covered by each dataset varied from ~100 mm² to ~700 mm². The total number of FOVs, unique sample areas, as well as the number of defocused images at each FOV used for training, validation and blind testing of the networks are summarized in Table S1.

### Network structure, training and validation

A GAN is used to perform Deep-R autofocusing (see Fig. S9). The GAN consists of a generator network and a discriminator network. The generator network follows a U-net[49] structure, consisting of five downsampling blocks with residual connections and five upsampling blocks. The discriminator network is a convolutional neural network, following a structure demonstrated in previous papers[24,32]. A discriminator loss based on high-level image features and structural loss terms calculated between the output and the spatially-registered ground truth images provide the objective function for training the generator. During the training phase, the network iteratively minimizes the loss functions of the generator and discriminator networks, defined as:

$$L_G = \lambda \times [1 - D(G(x))]^2 + \nu \times [1 - MSSSIM(y, G(x))] + \xi \times Berhu(y, G(x)) \quad (1)$$

$$L_D = D(G(x))^2 + [1 - D(y)]^2 \quad (2)$$

where $x$ represents the defocused input image, $y$ denotes the in-focus image used as ground truth, $G(x)$ denotes the generator output, $D(\cdot)$ is the discriminator inference. The generator loss function ($L_G$) is a combination the adversarial loss with two additional regularization terms: the multiscale structural similarity (MSSSIM) index[50] and the reversed Huber loss (BerHu)[51,52], balanced by regularization parameters $\lambda, \nu, \xi$. In our training, these parameters are set empirically such that three sub-types of losses contributed approximately equally after the convergence. MSSSIM is defined as:

$$MSSSIM(x, y) = \left[\frac{2\mu_{x_M}\mu_{y_M} + C_1}{\mu_{x_M}^2 + \mu_{y_M}^2 + C_1}\right]^{\alpha_M} \cdot \prod_{j=1}^{M}\left[\frac{2\sigma_{x_j}\sigma_{y_j} + C_2}{\sigma_{x_j}^2 + \sigma_{y_j}^2 + C_2}\right]^{\beta_j}\left[\frac{\sigma_{x_jy_j} + C_3}{\sigma_{x_j}\sigma_{y_j} + C_3}\right]^{\gamma_j} \quad (3)$$

where $x_j$ and $y_j$ are the distorted and reference images downsampled $2^{j-1}$ times, respectively; $\mu_x, \mu_y$ are the averages of $x, y$; $\sigma_x^2, \sigma_y^2$ are the variances of $x, y$; $\sigma_{xy}$ is the covariance of $x, y$; $C_1, C_2, C_3$ are constants used to stabilize the division with a small denominator; and $\alpha_M, \beta_j, \gamma_j$ are exponents used to adjust the relative importance of different components. The MSSSIM function is implemented using the TensorFlow[53] function *tf.image.ssim_multiscale*, using its default parameter settings. The BerHu loss is defined as:

$$Berhu(x, y) = \sum_{\substack{m,n \\ |x(m,n)-y(m,n)|\leq c}} |x(m,n) - y(m,n)| + \sum_{\substack{m,n \\ |x(m,n)-y(m,n)|> c}} \frac{[x(m,n) - y(m,n)]^2 + c^2}{2c} \quad (4)$$

where $x(m, n)$ refers to the pixel intensity at point *(m, n)* of an image of size $M \times N$, $c$ is a hyperparameter, empirically set as ~10% of the standard deviation of the normalized ground truth image. MSSSIM provides a multi-scale, perceptually-motivated evaluation metric between the generated image and the ground truth image, while BerHu loss penalizes pixel-wise errors, and assigns higher weights to larger losses exceeding a user-defined threshold. In general, the combination of a regional or a global perceptual loss, e.g., SSIM or MSSSIM, with a pixel-wise loss, e.g., L1, L2, Huber and BerHu, can be used as a structural loss to improve the network performance in image restoration related tasks[54]. The introduction of the discriminator helps the network output images to be sharper.

All the weights of the convolutional layers were initialized using a truncated normal distribution (Glorot initializer), while the weights for the fully connected (FC) layers were initialized to 0.1. An adaptive moment estimation (Adam)[55] optimizer was used to update the learnable parameters, with a learning rate of $5\times10^{-4}$ for the generator and $1\times10^{-6}$ for the discriminator, respectively. In addition, six updates of the generator loss and three updates of the discriminator loss are performed at each iteration to maintain a balance between the two networks. We used a batch size of 5 in our training phase, and the validation set was tested every 50 iterations. The training process converges after ~100,000 iterations (equivalent to ~50 epochs) and the best model is chosen as the one with the smallest BerHu loss on the validation set, which was empirically found to perform better. The details of the training and the evolution of the loss term are presented in Fig. S8. For each dataset with a different type of sample and a different imaging system, Deep-R network was trained from scratch.

For the optimization of the DPM decoder (Fig. 7(a)), the same structure of the up-sampling path of Deep-R network is used, and then optimized using an Adam optimizer with learning rate of $1 \times 10^{-4}$ and an L2-based objective function ($L_{Dec}$), as expressed below:

$$L_{Dec} = \sum_{m,n}(x(m,n) - y(m,n))^2$$

where $x$ and $y$ denote the output DPM and the ground-truth DPM, respectively, and $m, n$ stand for the lateral coordinates.

**Implementation details**

The network is implemented using TensorFlow on a PC with Intel Xeon Core W-2195 CPU at 2.3GHz and 256 GB RAM, using Nvidia GeForce RTX 2080Ti GPU. The training phase using ~30,000 image pairs (512×512 pixels in each image) takes about ~30 hours. After the training, the blind inference (autofocusing) process on a 512×512-pixel input image takes ~ 0.1 sec.

**Image quality analysis**

*Difference image calculation:* the raw inputs and the network outputs were originally 16-bit. For demonstration, we normalized all the inputs, outputs and ground truth images to the same scale. The absolute difference images of the input and output with respect to the ground truth were normalized to another scale such that the maximum error was 255.

*Image sharpness coefficient for tilted sample images:* Since there was no ground truth for the tilted samples, a reference image was synthesized using a maximum intensity projection (MIP) along the axial direction, incorporating 10 planes between z = 0 μm and z = 1.8 μm for the best visual sharpness. Following this, the input and output images were first convolved with a Sobel operator to calculate a sharpness map, *S*, defined as:

$$S(I) = \sqrt{I_X^2 + I_Y^2} \quad (5)$$

where $I_X, I_Y$ represent the gradients of the image *I* along X and Y axis, respectively. The relative sharpness of each row with respect to the reference image was calculated as the ordinary least square (OLS) coefficient without intercept[56]:

$$\hat{\alpha}_i = \frac{S(x)_i S(y)_i^T}{S(y)_i S(y)_i^T}, i = 1, 2, 3, \cdots, N \quad (6)$$

where $S_i$ is the i-th row of $S$, $y$ is the reference image, $N$ is the total number of rows.

The standard deviation of the relative sharpness is calculated as:

$$Std(\hat{\alpha}_i) = \sqrt{\frac{RSS_i}{(N-1)S(y)_i S(y)_i^T}} \quad (7)$$

$$RSS_i = \sum_i (S(x)_i - \hat{\alpha}_i S(y)_i)^2 \quad (8)$$

where $RSS_i$ stands for the sum of squared residuals of OLS regression at the i[th] row.

**Estimation of the lateral FWHM values for PSF analysis**

A threshold was applied to the most focused plane (with the largest image standard deviation) within an acquired axial image stack to extract the connected components. Individual regions of 30×30 pixels were cropped around the centroid of the sub-regions. A 2D Gaussian fit (*lsqcurvefit*) using MATLAB (MathWorks) was performed on each plane in each of the regions to retrieve the evolution of the lateral FWHM, which was calculated as the mean FWHM of x and y directions. For *each* of the sub-regions, the fitted centroid at the most focused plane was used to crop a x-z slice, and another 2D Gaussian fit was performed on the slide to estimate the axial FHWM. Using the statistics of the *input* lateral and axial FWHM at the focused plane, a threshold was performed on the sub-regions to exclude any dirt and bead clusters from this PSF analysis.

**Implementation of RL and Landweber image deconvolution algorithms**

The image deconvolution (which was used to compare the performance of Deep-R) was performed using the ImageJ plugin DeconvolutionLab2[41]. We adjusted the parameters for RL and Landweber algorithm such that the reconstructed images had the best visual quality. For Landweber deconvolution, we used 100 iterations with a gradient descent step size of 0.1. For RL deconvolution, the best image was obtained at the 100[th] iteration. Since the deconvolution results exhibit known boundary artifacts[57] at the edges, we cropped 10 pixels at each image edge when calculating the SSIM and RMSE index to provide a fair comparison against Deep-R results.

**Speed measurement of online autofocusing algorithms**

The autofocusing speed measurement is performed using the same microscope (IX83, Olympus) with a 20×/0.75NA objective lens using nanobead samples. The online algorithmic autofocusing procedure is controlled by the OughtaFocus plugin[44] in Micro-Manager[45], which uses the Brent's algorithm[46]. We choose the following search parameters: SearchRange = 10 μm, tolerance = 0.1 μm, exposure = 100 ms. Then, we compared the autofocusing time of 4 different focusing criteria: Vollath-4 (VOL4)[39], Vollath-5 (VOL5)[39], standard deviation (STD) and normalized variance (NVAR)[10]. These criteria are defined as follows:

$$F_{VOL4} = \sum_{m=1}^{M-1} \sum_{n=1}^{N} x(m,n)x(m+1,n) - \sum_{m=1}^{M-2} \sum_{n=1}^{N} x(m,n)x(m+2,n) \quad (9)$$

$$F_{VOL5} = \sum_{m=1}^{M-1} \sum_{n=1}^{N} x(m,n)x(m+1,n) - MN\mu^2 \quad (10)$$

$$F_{STD} = \sqrt{\frac{1}{MN} \sum_{m=1}^{M} \sum_{n=1}^{N} [x(m,n) - \mu]^2} \quad (11)$$

$$F_{NVAR} = \frac{1}{MN\mu} \sum_{m=1}^{M} \sum_{n=1}^{N} [x(m,n) - \mu]^2 \quad (12)$$

where $\mu$ is the mean intensity defined as:

$$\mu = \sum_{m=1}^{M} \sum_{n=1}^{N} x(m,n) \quad (13)$$

The autofocusing time is measured by the controller software, and the exposure time for the final image capture is excluded from this measurement. The measurement is performed on 4 different FOVs, each measured 4 times, with the starting plane randomly initiated from different heights. The final statistical analysis (Table S2) was performed based on these 16 measurements.

# Supplementary Information for
# Single-shot autofocusing of microscopy images using deep learning

Yilin Luo[1,2,3]†, Luzhe Huang[1,2,3]†, Yair Rivenson[1,2,3]*, Aydogan Ozcan[1,2,3,4]*

1 Electrical and Computer Engineering Department, University of California, Los Angeles, California 90095, USA

2 Bioengineering Department, University of California, Los Angeles, California 90095, USA

3 California Nano Systems Institute (CNSI), University of California, Los Angeles, California 90095, USA

4 David Geffen School of Medicine, University of California Los Angeles, California 90095, USA

* ozcan@ucla.edu, rivensonyair@ucla.edu

† Contributed equally

**Content of Supplementary Information:**

    10 pages, 9 figures and 2 tables

## S. 1 Deep-R based autofocusing of non-uniformly defocused samples

Although Deep-R is trained on uniformly defocused microscopy images, during blind testing it can also successfully autofocus non-uniformly defocused images without prior knowledge of the image distortion or defocusing (see e.g., Fig. 7 of the main text). As another example, Fig. S1 illustrates Deep-R based autofocusing of a non-uniformly defocused image of a human breast tissue sample that had ~1.5° planar tilt (corresponding to an axial difference of $\delta z$ = 4.356 µm within the effective FOV of a 20×/0.75NA objective lens). This Deep-R network was trained using only uniformly defocused images and is the same network that generated the results reported in Fig. 3. As illustrated in Fig. S1, at different focal depths (e.g., $z$ = 0 µm and $z$ = -2.2 µm), because of the sample tilt, different sub-regions within the FOV were defocused by different amounts, but they were simultaneously autofocused by Deep-R, all in parallel, generating an extended DOF image that matches the reference image (Fig. S1(b), see the Methods section). Moreover, we quantified the focusing performance of Deep-R on this tilted sample using a row-based sharpness coefficient (Fig. S1(c), see the Methods section), which reports, row by row, the relative sharpness of the output (or the input) images with respect to the reference image along the direction of the sample tilt (i.e., y-axis). As demonstrated in Fig. S1(c), Deep-R output images achieved a significant increase in sharpness measure within the entire FOV, validating Deep-R's autofocusing capability for a non-uniformly defocused, tilted sample. Fig. S1(c) was calculated on a single sample FOV; Fig. S1(d) reports the statistical analysis of Deep-R results on the whole testing image dataset consisting of 18 FOVs that are each non-uniformly defocused, confirming the same conclusion as in Fig. S1(c).

## S. 2 Comparison between Deep-R and Deep-Z

In comparison to our earlier work, Deep-Z, which requires a user-defined DPM for on demand refocusing of a fluorescence image to a desired surface, Deep-R achieves blind autofocusing using a single acquired image without any prior knowledge of the defocus amount, direction, or the aberration pattern. Stated differently, Deep-Z can be analogous to the focusing knob of a microscope that is digitally operated by a user, and Deep-R is the blind autofocusing unit that automatically finds the best focused image in a single shot without any user interventions or a priori information of the defocusing distortion. To quantitatively compare Deep-R with Deep-Z and have a better understanding of their unique differences, we used 300 nm fluorescence beads that were prepared and imaged in the same way as in Fig. 4 of main text. The two deep networks (Deep-R and Deep-Z) were trained on the same dataset with the same defocusing range (±10µm) and axial scanning step size (0.2 µm). As shown in Fig. S2(a), we used an image at the axial position of z=-2 µm as the input image (defocused), and generated a *virtual image stack* using Deep-Z over an axial range of z=-2 µm to z=2 µm with Δz=0.2 µm intervals. The resulting virtually propagated Deep-Z images, along with the original input image at z=-2 µm were then all passed through Deep-R for autofocusing. Figure S2(b) compares the lateral FWHM values of all the nanobeads measured in Deep-Z and Deep-R output images. This comparative analysis and the results highlight that Deep-R holds some unique features for image autofocusing.

First, Deep-R provides an improvement in term of the overall focus quality as shown in the lateral FWHM mean values reported in Fig. S2(b). Stated differently, a search of the correct focus within Deep-Z generated virtual image stack is inferior to the Deep-R autofocusing result that is achieved in a single shot. This is expected as Deep-Z is not trained for bringing all the features/objects within the input FOV into focus, in fact, it is trained to perform virtual propagation of input fluorescence images to a different surface defined by a desired DPM. Deep-R, on the other hand, is trained to only perform autofocusing, and virtually increases the depth of field of the imaging system.

Second, Deep-R provides a global autofocusing performance in one step blind inference, which is evident by the reduced standard variance of the lateral FWHM values (light blue shadow) compared to the larger standard variance observed in Deep-Z output images. This is related to the fact that Deep-R can autofocus $n$ different objects at $m$ different axial positions within its training range using only one step inference (i.e., with a computational time complexity of $O(1)$), rather than requiring $O(mn)$ steps that Deep-Z would need to perform for a non-uniform DPM search in order to refocus different objects within the ROI.

Third, as demonstrated in Fig. S2, Deep-R can also be used to complement the results generated by Deep-Z. Both the mean and variance of the FWHM values corresponding to the nanobeads within the sample FOV are improved by Deep-R based autofocusing performed on each one of the Deep-Z output images.

**Supplementary Figures and Tables**

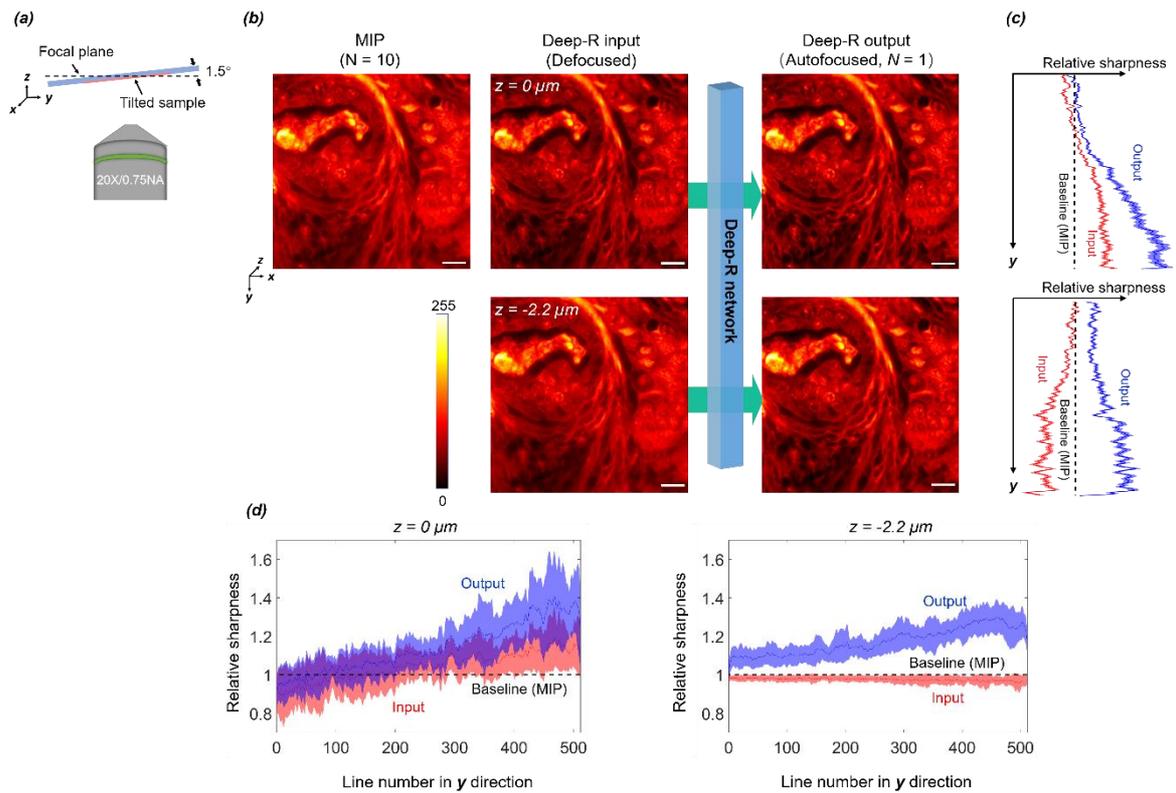

**Fig. S1. Deep-R blind autofocusing of non-uniformly defocused fluorescence images.** (a) Image acquisition of a tilted autofluorescent sample, corresponding to a depth difference of $\delta z$ = 4.356 µm within the FOV. (b) Deep-R autofocusing results for a tilted sample. Since no real ground truth is available, we used the MIP image, calculated from $N$ = 10 images as the reference image in this case. Top row: autofocusing of an input image where the upper region is blurred due to the sample tilt. Second row: autofocusing of an input image where the lower region is blurred due to the sample tilt. Scale bars, 20 µm. (c) Deep-R output images are quantitatively evaluated using a relative sharpness coefficient that compares the sharpness of each row with the baseline (MIP) image as well as the input image shown in (b). (d) The statistics were calculated from a testing dataset containing 18 FOVs, each with 512×512 pixels.

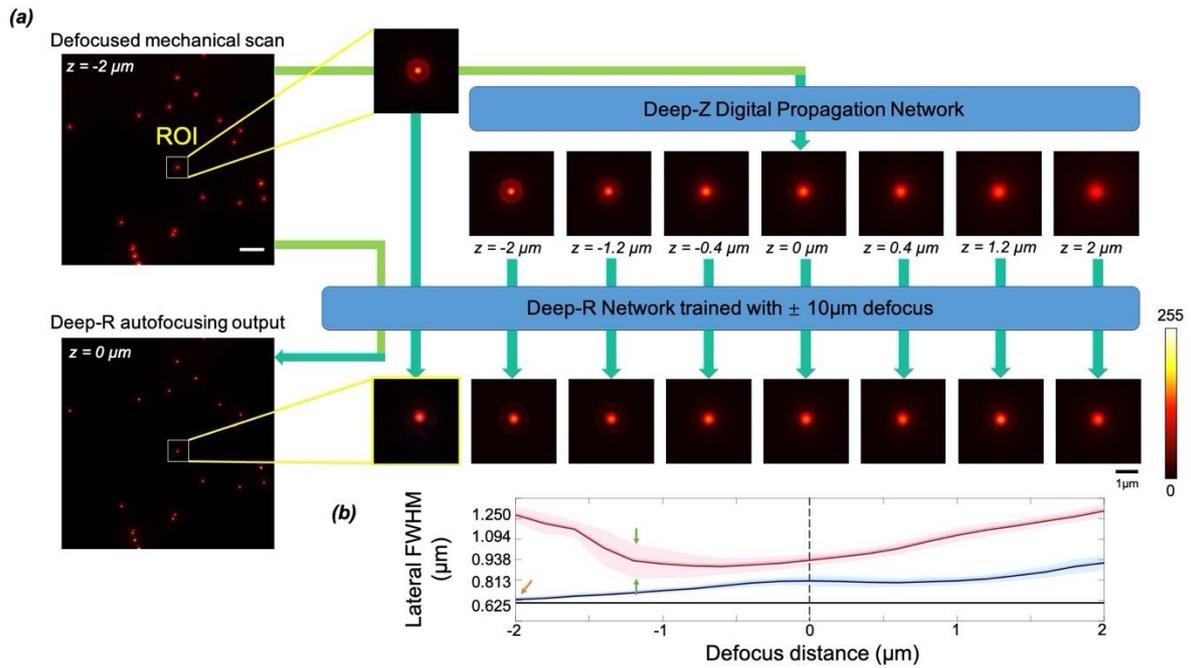

**Fig. S2. Comparison between Deep-Z and Deep-R.** (a) We used a defocused input image (z=-2 μm) of 300 nm fluorescence beads, and generated a virtual image stack using Deep-Z over an axial range of z=-2 μm to z=2 μm with Δz=0.2 μm intervals and uniform DPMs. The resulting virtually propagated Deep-Z images, along with the original input image at z=-2 μm were all passed through Deep-R for autofocusing. Scale bar: 5μm. (b) The lateral FWHM values of the nanobeads. Red: Deep-Z output images for each axial distance; Blue: Deep-R autofocused images; Black (baseline): Deep-R based autofocused image on the original input image at z=-2 μm; Dashed line: the correct in-focus position; Orange arrow: the small FWHM gap is caused by the differences between the corresponding Deep-R inputs: the original input image at z=-2 μm vs. the Deep-Z digital propagation result with a uniform DPM of Δz=0 μm. The solid lines indicate the mean FWHM values and the shadows represent the standard variance from individual nanobeads ($N \approx 20$).

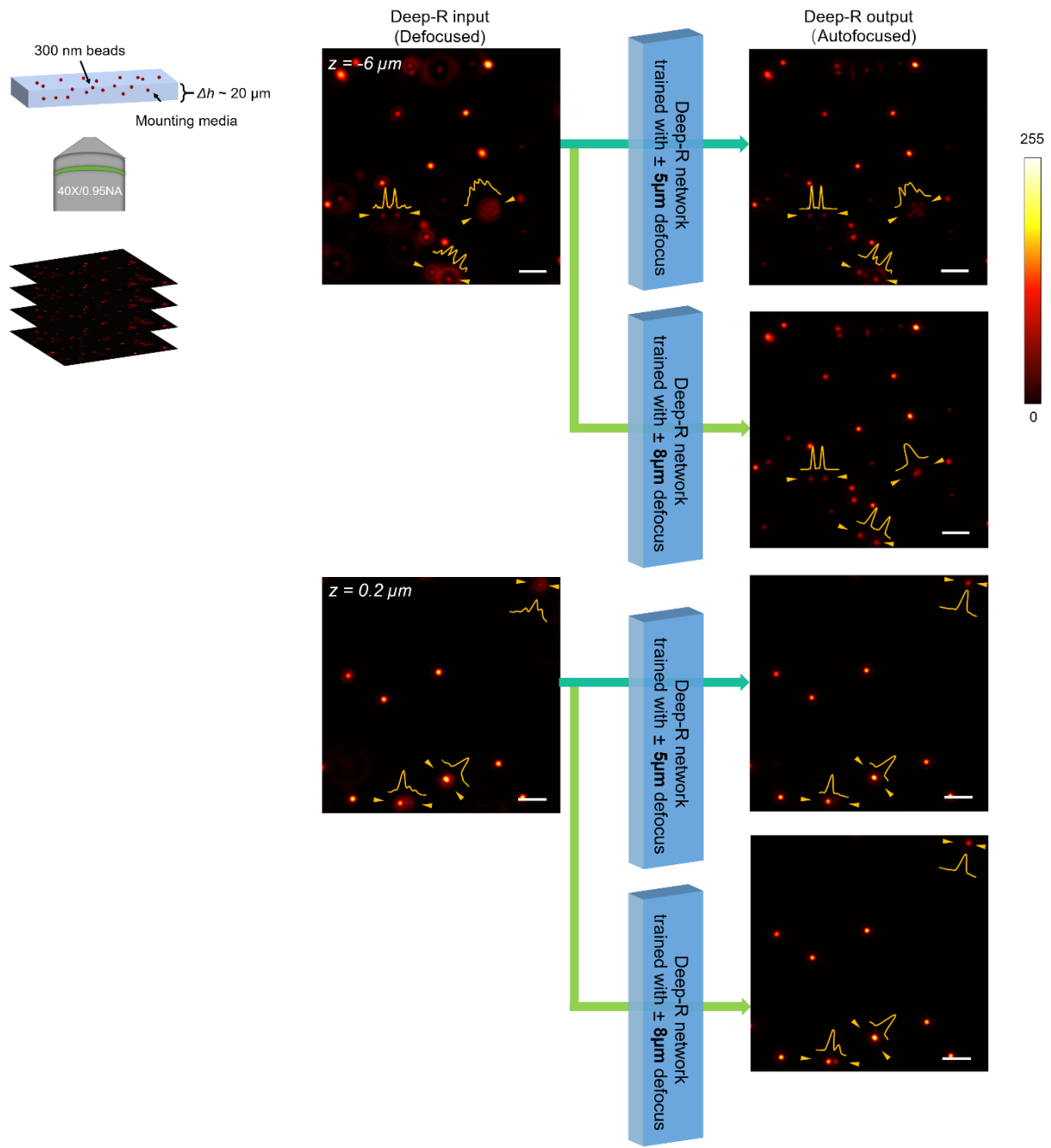

**Fig. S3. Deep-R based autofocusing of a sample with nanobeads dispersed in 3D.** 300nm beads are randomly distributed in a sample volume of ∼ 20 μm thickness. Using a Deep-R network trained with ±5 μm defocus range, autofocusing on some of these nanobeads failed since they were out of this range. These beads, however, were successfully refocused using a network trained with ±8 μm defocus range. Scale bar: 5 μm.

S5

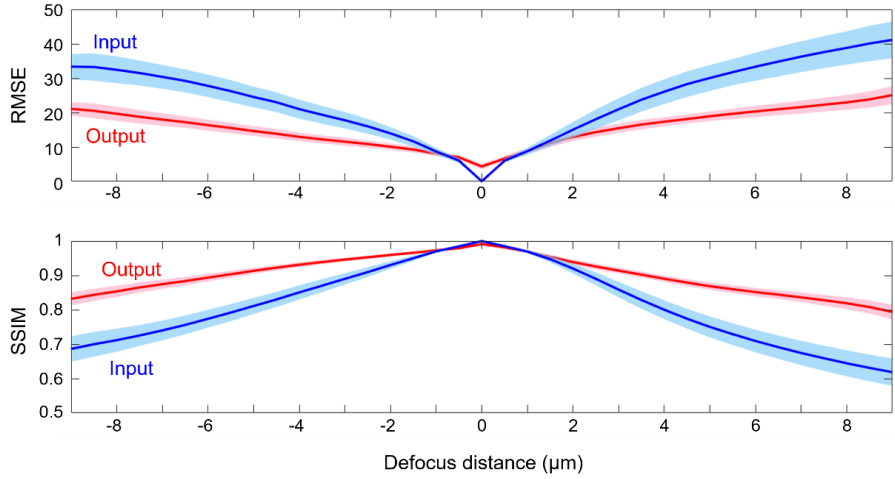

**Fig. S4. Deep-R blind autofocusing performance on brightfield microscopy images.** Mean and standard deviation of SSIM and RMSE values of the input and output images with respect to the ground truth ($z = 0$ µm, in-focus image) are plotted as a function of the axial defocus distance. The statistics are calculated from a testing dataset containing 58 FOVs, each with 512×512 pixels.

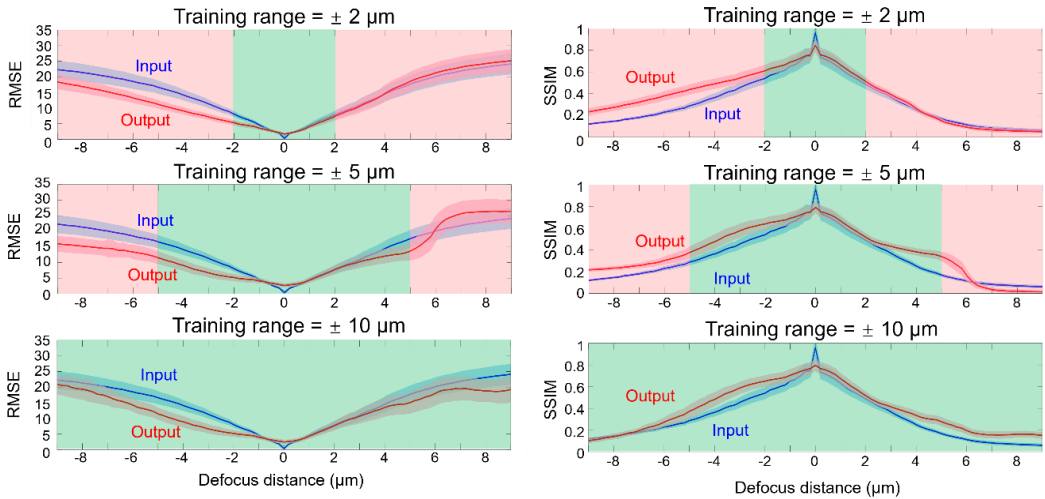

**Fig. S5. Comparison of Deep-R autofocusing performance using different defocus training ranges.** Mean and standard deviation of (a) RMSE and (b) SSIM values of the input and output images at different defocus distances. Three different Deep-R networks are reported here, each trained with a different defocus range, spanning ± 2µm, ± 5µm, and ± 10µm, respectively. The curves are calculated using 26 unique sample FOVs, each with 512×512 pixels.

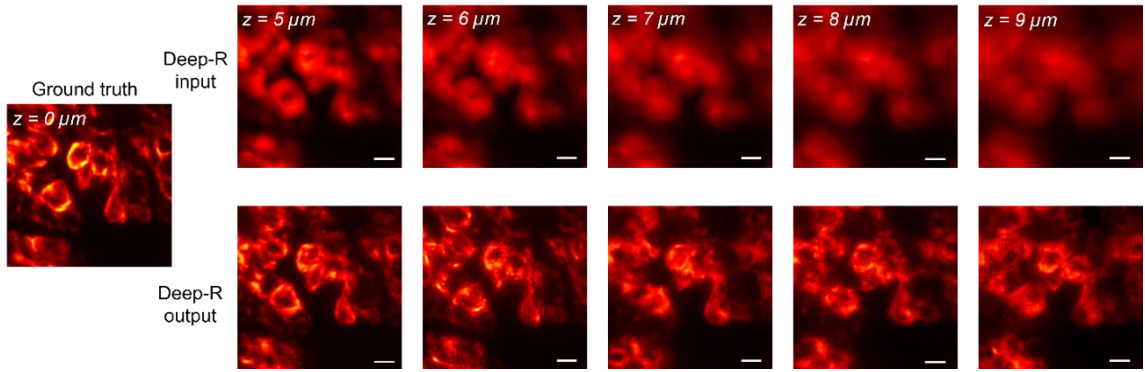

**Fig. S6. Deep-R based blind autofocusing of images captured at large defocus distances. Scale bar: 10 μm.**

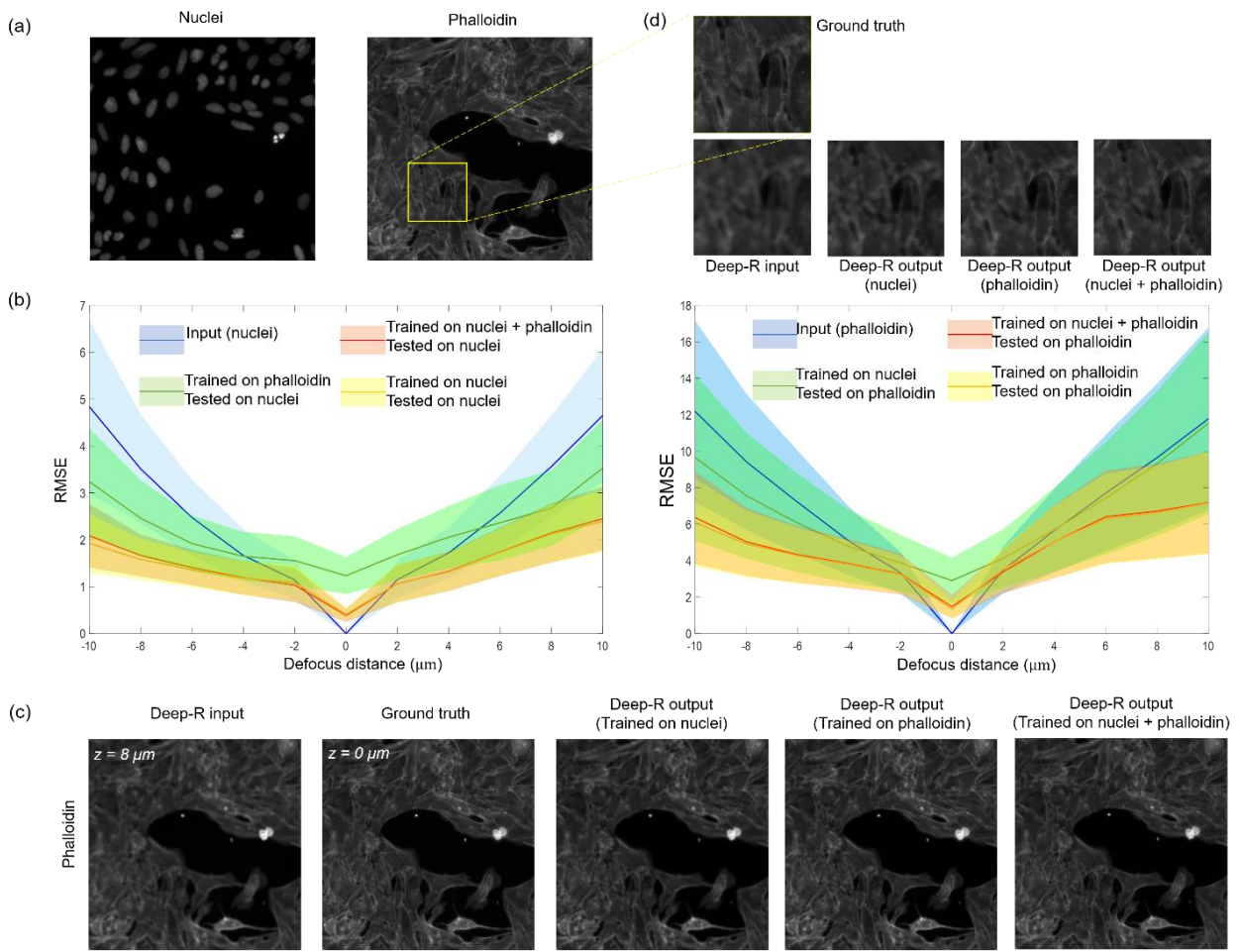

**Fig. S7 Deep-R generalization to new sample types.** We separately trained 3 Deep-R networks with a defocus range of ±10 μm on 3 different datasets that contain images of only nuclei, only phalloidin and both types of images. The networks were then blindly tested on different types of samples. (a) Sample images of nuclei and phalloidin. (b) The input and output of the three networks are compared under the RMSE value with respect to the ground truth ($z = 0$ μm). Blue curve: network input. Green Curve: output from the network that *did not* train on the type of sample. Red curve: output from the network trained with a *mixed* type of samples. Yellow curve: output from the network trained *with* the type of sample. (c) A model trained with nuclei images brings back some details when tested on phalloidin images. However, the autofocusing is not optimal, compared with the reconstruction using a model that was trained only with phalloidin images. (d) Zoomed-in regions of the ground truth, input and Deep-R output images in (c). The yellow frame highlights the selected region.

S7

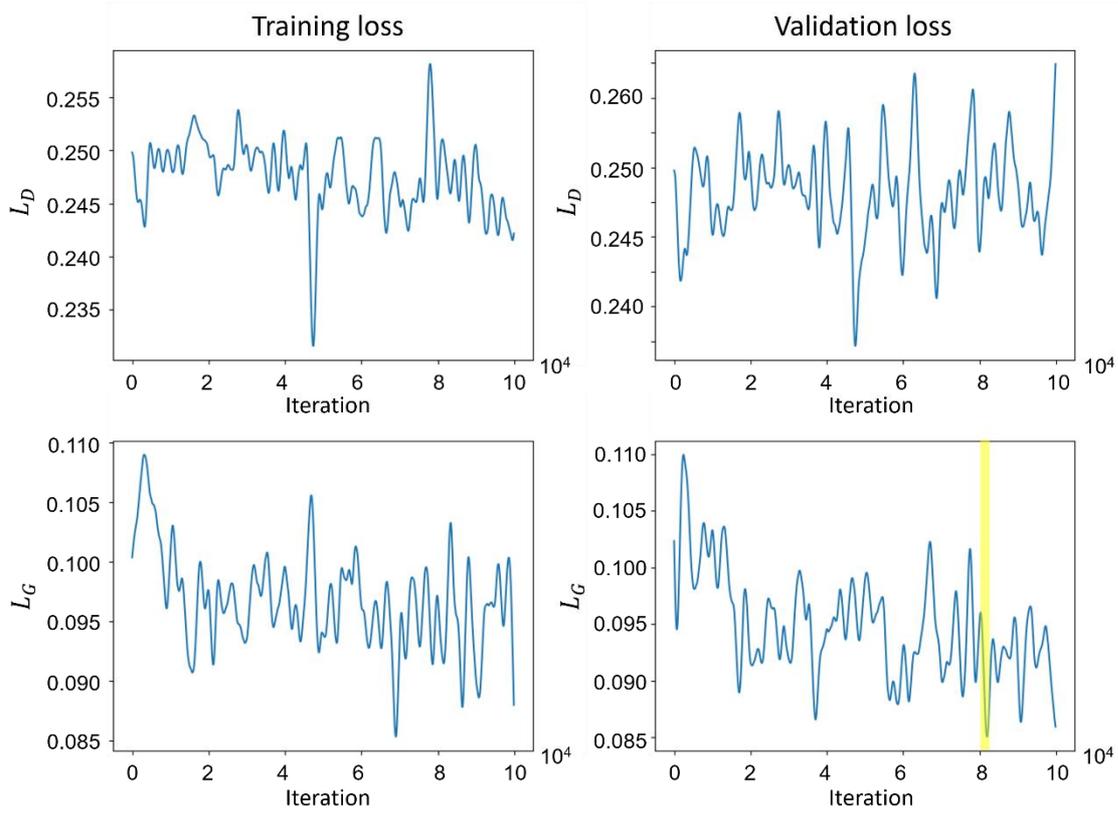

**Fig. S8. The training and validation loss curves as a function of the training iterations.** Deep-R was trained from scratch on breast tissue sample dataset. For easier visualization, the loss curves are smoothed using a Hanning window of size 1200. Due to the least square form of the discriminator loss, the equilibrium is reached when $L_D \approx$ 0.25. Optimal model was reached at ~80K iterations (yellow highlighted).

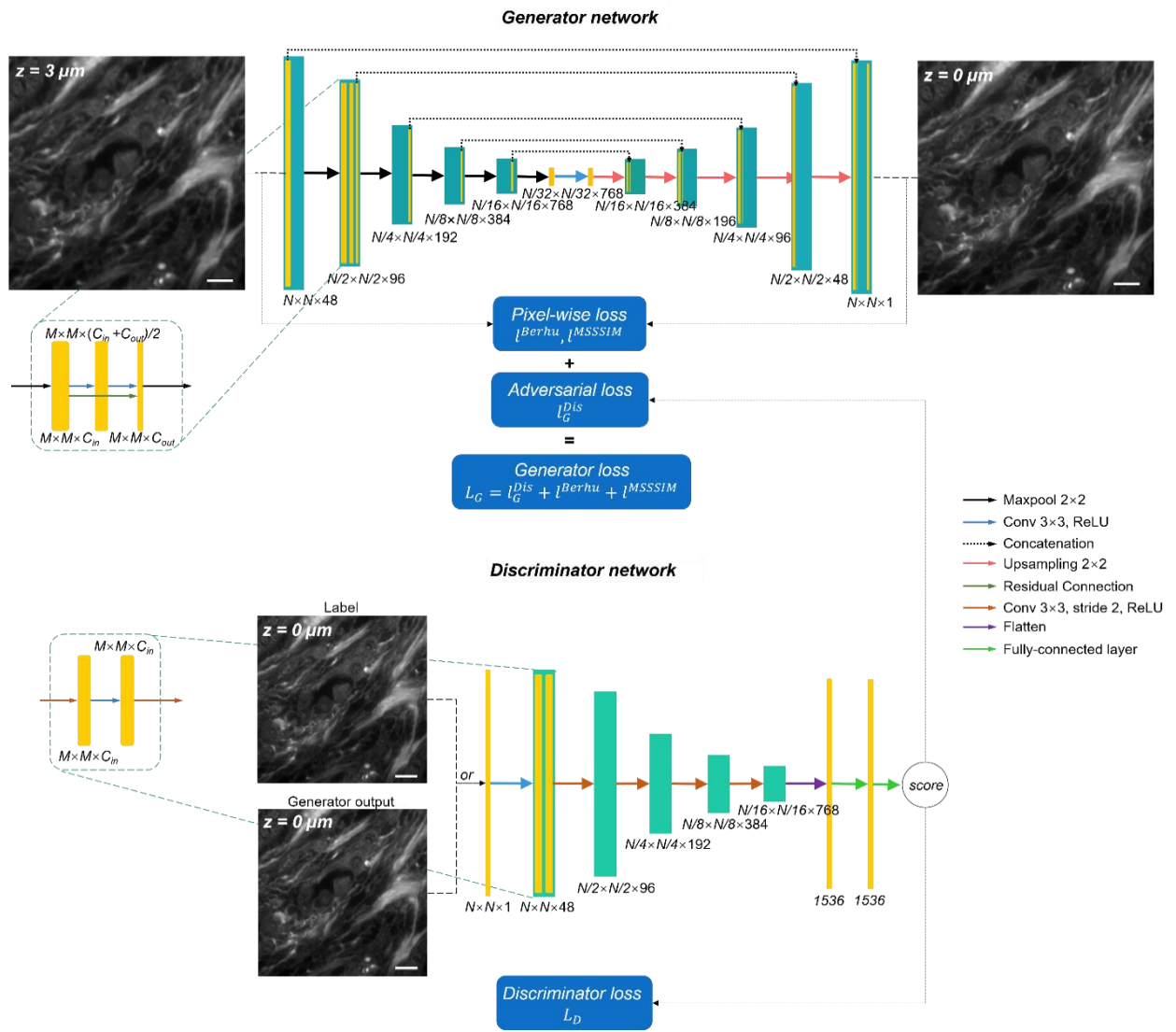

**Fig. S9. Deep-R network architecture.**

|  | Training set (FOV) | Validation set (FOV) | Testing set (FOV) | Unique sample area (mm$^2$) | Training defocus range (μm) | z step size (μm) | Depths at each FOV |
|---|---|---|---|---|---|---|---|
| Flat breast tissue (20X, DAPI) | 1156 | 118 | 6 | 446 | ±5 μm | 0.5 | 21 |
| Tilted breast tissue (20X, DAPI) | / | / | 18 | 6.3 | / | 0.2 |  |
| Ovary tissue (40X, Cy5) | 874 | 218 | 26 | 97 | ±2 μm, ±5 μm, ±10 μm | 0.2 | 21, 51, 101 |
| H&E Stained prostate (20X, Brightfield) | 1776 | 205 | 58 | 710 | ±10 μm | 0.5 | 41 |
| 300nm fluorescent beads (40X, Texas Red) | 1077 | 202 | 20 | 113 | ±5 μm, ±8 μm | 0.2 | 51, 81 |
| Human U2OS cells (20X, two channels for nuclei and phalloidin, respectively) | 345 for each channel | 51 for each channel | 38 for each channel | / | ±10 μm | 2 | 11 |

Table. S1. Distribution of the training, validation and testing image datasets. At each FOV, a *z*-stack that covers a certain axial defocus range is obtained, where each image in the *z*-stack has a size of 512×512 pixels.

| Focusing criterion | Average time (sec/mm$^2$) | Standard deviation (sec/mm$^2$) |
|---|---|---|
| Vollath4[39] | 42.91 | 3.16 |
| Vollath5[39] | 39.57 | 3.16 |
| Standard deviation | 37.22 | 3.07 |
| Normalized variance | 36.50 | 0.36 |
| Deep-R (CPU) | 20.04 | 0.23 |
| Deep-R (GPU) | 2.98 | 0.08 |

Table S2. Comparison of Deep-R computation time per 1 mm$^2$ of sample FOV (captured using a 20×/0.75NA objective lens) compared against other state-of-the-art online autofocusing methods.